\input{aipcheck}

\documentclass[
    ,final            
  ]
  {aipproc}

\layoutstyle{8x11double}

\newcommand{\ra}{\rangle}
\newcommand{\la}{\langle}

\newcommand{\cM}{{\cal M}}

\newcommand{\cT}{{\cal T}}

\begin{document}

\title{The transfer matrix in four dimensional causal dynamical triangulations}

\classification{12.10.-g, 14.70.Kv, 04.60.-m, 04.60.Gw, 04.60.NC, 05.10.Ln}
\keywords      {quantum gravity, nonperturbative quantization, causal dynamical triangulations, dynamical triangulations, lattice gravity}

\author{J. Ambj\o rn}{
  address={The Niels Bohr Institute, Copenhagen University,
Blegdamsvej 17, DK-2100 Copenhagen \O , Denmark.}
  ,altaddress={Radboud University, Nijmegen, Institute for Mathematics, Astrophysics and Particle Physics, Heyendaalseweg 135, 6525 AJ Nijmegen, The Netherlands}
}

\author{ J. Gizbert-Studnicki}{
  address={Institute of Physics, Jagiellonian University,
Reymonta 4, PL 30-059 Krakow, Poland}
}

\author{ A. T. G\"{o}rlich}{
  address={Institute of Physics, Jagiellonian University,
Reymonta 4, PL 30-059 Krakow, Poland}
  ,altaddress={The Niels Bohr Institute, Copenhagen University,
Blegdamsvej 17, DK-2100 Copenhagen \O , Denmark.} 
}

 \author{ J. Jurkiewicz}{
  address={Institute of Physics, Jagiellonian University,
Reymonta 4, PL 30-059 Krakow, Poland}
}

 \author{R. Loll}{
  address={Radboud University, Nijmegen, Institute for Mathematics, Astrophysics and Particle Physics, Heyendaalseweg 135, 6525 AJ Nijmegen, The Netherlands}
}

\begin{abstract}

The Causal Dynamical Triangulation model of quantum gravity (CDT)
is a proposition to evaluate the path integral over space-time geometries
using a lattice regularization with a discrete proper time and geometries realized as simplicial
manifolds. The model admits a Wick rotation to imaginary time for each space-time configuration.
Using computer simulations we determined the phase structure of the
model and discovered that it predicts a de Sitter phase with a four-dimensional
spherical semi-classical background geometry. 
The  model has a transfer matrix, relating spatial geometries at adjacent 
(discrete lattice) times. The transfer matrix uniquely determines
the theory. We show that the measurements of the scale factor
of the (CDT) universe are well described by an effective 
transfer matrix where the matrix elements are labelled only 
by the scale factor.  Using computer simulations we determine the 
effective transfer matrix elements and show how they 
relate to an effective minisuperspace action at all scales.

\end{abstract}

\maketitle


\section{Introduction}

The aim of the CDT approach is to evaluate the gravitational quantum amplitude
\begin{equation}
G({\bf g}_{\rm i},{\bf g}_{\rm f},t):=\;\;\;\sum\hspace{-1.15cm}
\int\limits_{ {\rm geometries:}\, {\bf g}_{\rm i}\rightarrow
{\bf g}_{\rm f}} {\cal D}[g]~{\rm e}^{iS[{\bf g_{\mu\nu}};t']}
\label{e0}
\end{equation}
 between  initial and final geometries $\{ {\bf g}_{\rm i},{\bf g}_{\rm f}\}$. In this version
 we do not include matter fields in the theory.
We use the intuition based on methods of Quantum Field Theory to view this amplitude as
a path integral over space-time geometries. Defining a path integral in this case requires
solving a number of non-trivial conceptual problems:
\begin{itemize}
\item Definition of {\em time evolution} assigns a special role to be played by  a proper
time $t$ for each quantum space-time, giving a meaning to the idea of {\em initial} and {\em final}
and introducing a time foliation of geometry.
\item At each time $t$ we should define a Hilbert space of states - spatial geometries of the Universe.
\item Definition of the measure ${\cal D}[g]$ is related to the choice of  {\em the domain of integration} (space of admissible space-times
we should include in the path integral) and possibly also solving the problem of  diffeomorphism invariance.
\end{itemize}

At the same time one would like to obtain an approach which is 
\begin{itemize}
\item Background independent - the background geometry may emerge dynamically, but should not be introduced a priori.
\item Non-perturbative - meaning again that it is not obtained as a perturbation around some fixed background.
\item Has a well-defined infrared limit, described by the General Theory of Relativity.
\end{itemize}
CDT provides a construction satisfying these requirements \cite{ajl1}. 
\begin{itemize}
\item We consider only space-time geometries which admit a global time foliation.
Causality means that the spatial topology of the Universe is fixed during the time evolution. 
\item We introduce a lattice regularization of geometries, assuming that both space and time are discretized. 
The time variable is indexed by an integer time. At a fixed time we construct
the Hilbert space of states  defined
as a set of states $|T \ra$ representing triangulations of a 3d topological  sphere using regular simplices (tetrahedra) with a common edge length $a_s$. 
This definition does not involve coordinates, being by construction diffeomorphism invariant. Different triangulations
$|T \ra$ correspond to different geometries, which cannot be mapped onto each other.
The states $|T \ra$  satisfy
\begin{equation}
\la T|T'\ra =\frac{1}{C_{T}}\delta_{T T'}
\label{e01}
\end{equation}
where $C_{T}$ is the order of the automorphism group of $T$. In the construction of the Hilbert space
we consider only three-manifolds with the simplest topology of of a three-sphere $S^3$. This space splits in a natural way
into a simple sum of spaces labelled by the number of tetrahedra $N$. The number of states for
a fixed volume $N$ is finite, but large. It grows exponentially with $N$ for large $N$.
\item Tetrahedra at time $t$ are bases of four-simplices $\{1,4\}$ and $\{4,1\}$ with four vertices
at time $t$ and one vertex at $t\pm 1$. To form a closed four-dimensional manifold we need also
four-simplices $\{3,2\}$ and $\{2,3\}$ with three vertices (triangle) at $t$ and two (spanning a link) at $t\pm 1$. Four-simplices
have a common length of the {\em time link} equal to $\tilde{a}_t$ {see Fig.\ \ref{Fig01}}.
\begin{figure}
\centerline{\scalebox{0.30}{\rotatebox{0}{\includegraphics{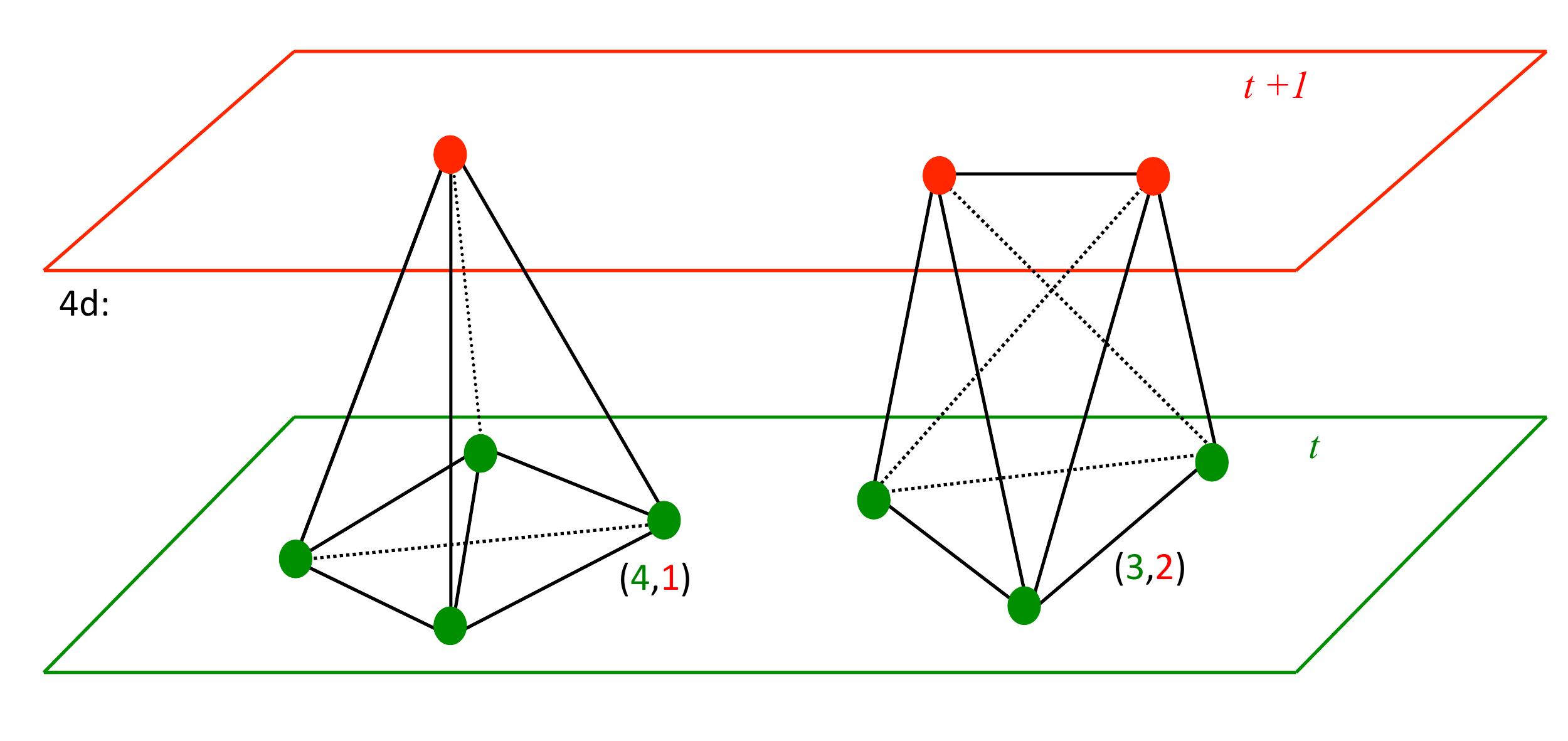}}}}
\caption{Building blocks of a simplicial four-manifold}
\label{Fig01}
\end{figure}
The sum (integral) over space-times is regularized as a sum over simplicial manifolds with topology
 $S^3 \times [0,1]$. 
Wick rotation to imaginary time can be realized as analytic continuation in $\tilde{a}_t \to {\rm i} a_t$ and can be performed
for each space-time configuration.
\end{itemize}
Manifolds can be characterized by a set of global numbers, where $N_4^{(4,1)}$ and $N_4^{(3,2)}$ denote the numbers of four-simplices
of a particular type and $N_0$ the number of vertices. Other numbers of this type can be expressed by this triple using
topological identities. After a Wick rotation each space-time configuration appears in the sum with the {\em real} weight
$\exp(-S(N_4^{(4,1)},N_4^{(3,2)},N_0))$, where
\begin{eqnarray}
S(N_4^{(4,1)},N_4^{(3,2)},N_0))=\Delta(2N_4^{(4,1)}+N_4^{(3,2)})\\ \nonumber
-(\kappa_0+6\Delta)N_0+\kappa_4(N_4^{(4,1)}+N_4^{(3,2)})
\label{action}
\end{eqnarray}
is the Hilbert-Einstein action calculated using the simplicial structure defined above. Dimensionless
coupling constants $\kappa_0$ and $\kappa_4$ are related respectively to the inverse gravitational constant
and cosmological constant. The parameter $\Delta$ is related to the ratio of the lattice spacings in time and spatial direction
($\Delta = 0$ for $a_s = a_t$). For the imaginary time the quantum amplitude has the form of a partition function
of a statistical ensemble of discretized space-time geometries. 

One can observe that the amplitude (partition function) can be represented as a {\em matrix}
product
\begin{eqnarray}
&&G(T_i,T_f; t) = \\ \nonumber 
&&\hspace{-0.8cm}\sum_{T_1,T_2,\cdots,T_{t-1}}
\langle T_0|{\cM}|T_1\rangle C(T_1)\langle T_1|{\cM}|T_2\rangle\cdots\cdots
\langle T_{t-1} |{\cM}|T_f\rangle
\label{quant}
\end{eqnarray}
Matrix elements $\langle T|{\cM}| T'\rangle $ depend on a number of  distinct
ways to connect geometric spatial states at times $t$ and $t+1$.

\section{Numerical results}

In practice the model cannot be solved analytically, except in the simplest case of 1+1 dimensional space-time \cite{al}.
For larger dimensionality we are forced to use numerical methods as a tool to obtain physical information about the system.
The tools we use are Monte Carlo simulations.
In the CDT model a very important ingredient of the theory becomes the entropy of configurations. 
For a fixed set of $N_0,N_4^{(4,1)},N_4^{(3,2)}$ the number of space-time configurations grows
exponentially like
\begin{equation}
{\cal N}(N_0,N_4^{(4,1)},N_4^{(3,2)}) \propto \exp(\kappa_4^c (N_4^{(4,1)}+N_4^{(3,2)})).
\label{e02}
\end{equation}
The critical parameter $\kappa_4^c$ depends on $\kappa_0$ and $\Delta$, which means
that the parameter $\Delta$, which originally had a geometric origin, plays the role of an
independent coupling constant. The critical parameter renormalizes the bare cosmological constant
$\kappa_4 \to \kappa_4^{eff}=\kappa_4 - \kappa_4^c$ and the model is defined
only for $\kappa_4^{eff} > 0$. In the limit $\kappa_4^{eff} \to 0^+$ the average
total number of simplices goes to infinity. This is the limit relevant for the continuum,
where we may discuss what happens when the discretization effects  can be neglected.

In practice, the numerical approach means that we must consider systems with a finite volume.
We can recover the information about the continuum properties studying the scaling properties of
observables for a sequence of large but finite $N_4 \to \infty$. This reduces the parameter space
of the model to a set of two bare coupling constants $\kappa_0$ and $\Delta$. In our simulations
we choose periodic boundary conditions, which on the one hand frees us from the necessity to define initial and final
geometries, but on the other hand does not change the physical picture, as will become clear below.

\subsection{Phase structure of CDT}

Using the Monte Carlo program  we perform a random walk in the space of configurations 
using 7 elementary local {\em moves}, which preserve the (local and global)  topological restrictions on a manifold. 
The probability
to perform a particular move is obtained by the detailed balance condition. This is a Markov process with a stationary limiting distribution
satisfying
\begin{equation}
Prob({\cT}) \propto \exp\left(-S({\cT})\right)
\label{e03}
\end{equation}
Configurations separated by a large number of moves are {\em statistically independent} and
the probability to obtain a particular configuration is given by the limiting distribution.
Expectation values of observables are measured as averages in the large but finite set of statistically independent space-time configurations
obtained at a particular set of parameters  $\kappa_0$ and $\Delta$ and (approximately) fixed  $N_4^{(4,1)}$.
The measurements are repeated for an increasing sequence of $N_4^{(4,1)}$ to check the scaling.
In the following we discuss a very useful observable characterizing each space-time configuration. It is the distribution 
of a three-volume $N(t)$ as a function of discrete time $t$.

Depending on the position in the $\{\kappa_0,\Delta\}$ plane our system appears to be in
three physically distinct phases \cite{ajl2} characterized by different behaviour of $N(t)$.
\begin{figure*}
\centerline{\scalebox{0.3}{\rotatebox{0}{\includegraphics{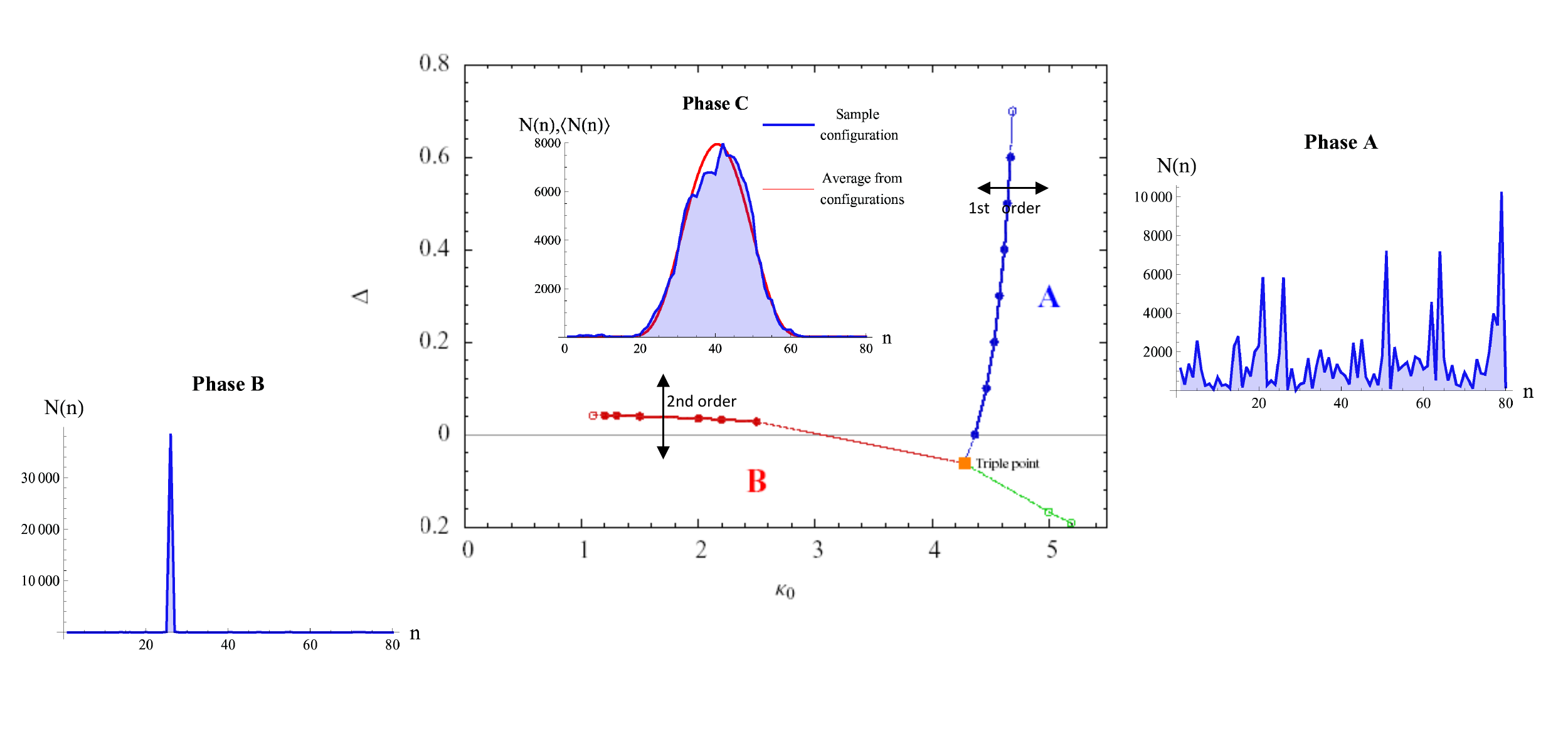}}}}
\caption{Three phases of the four-dimensional CDT.}
\label{Fig02}
\end{figure*}
On the Fig.\ \ref{Fig02} we show a sample distribution of the three-volume $N(t)$ for one
typical configuration in phases A, B and C. 
Phase A is characterized be a sequence of slices $N(t)$ with no correlation between 
the states at neighbouring times.
In phase B  the time dependence of the distribution is squeezed
to one time value (one may view it as a spontaneous compactification of the time variable).
For other times the volume is close to minimal. It cannot be completely zero, because
we choose periodic boundary conditions and do not allow the volume to vanish at any fixed time $t$.
Most interesting from a physical point of view is the C phase. The 
volume profile looks like a fluctuation superimposed over a regular classical background (the red line on the plot).
A typical configuration consists of a central {\em blob} and a {\em stalk} of cut-off size
resulting again from our choice of boundary conditions (periodicity in time).
The red line is the average distribution over many configurations with the same volume.
We can compare distributions for a sequence of volumes (Fig.\ \ref{fix}) and we find a universal scaling
behaviour in the variable  $\bar{t} = (n-n_c)/N_4^{(1/d_H)}$, with Hausdorff dimension $d_H \approx 4$.
\begin{figure}[b]
\centerline{\scalebox{0.3}{\rotatebox{0}{\includegraphics{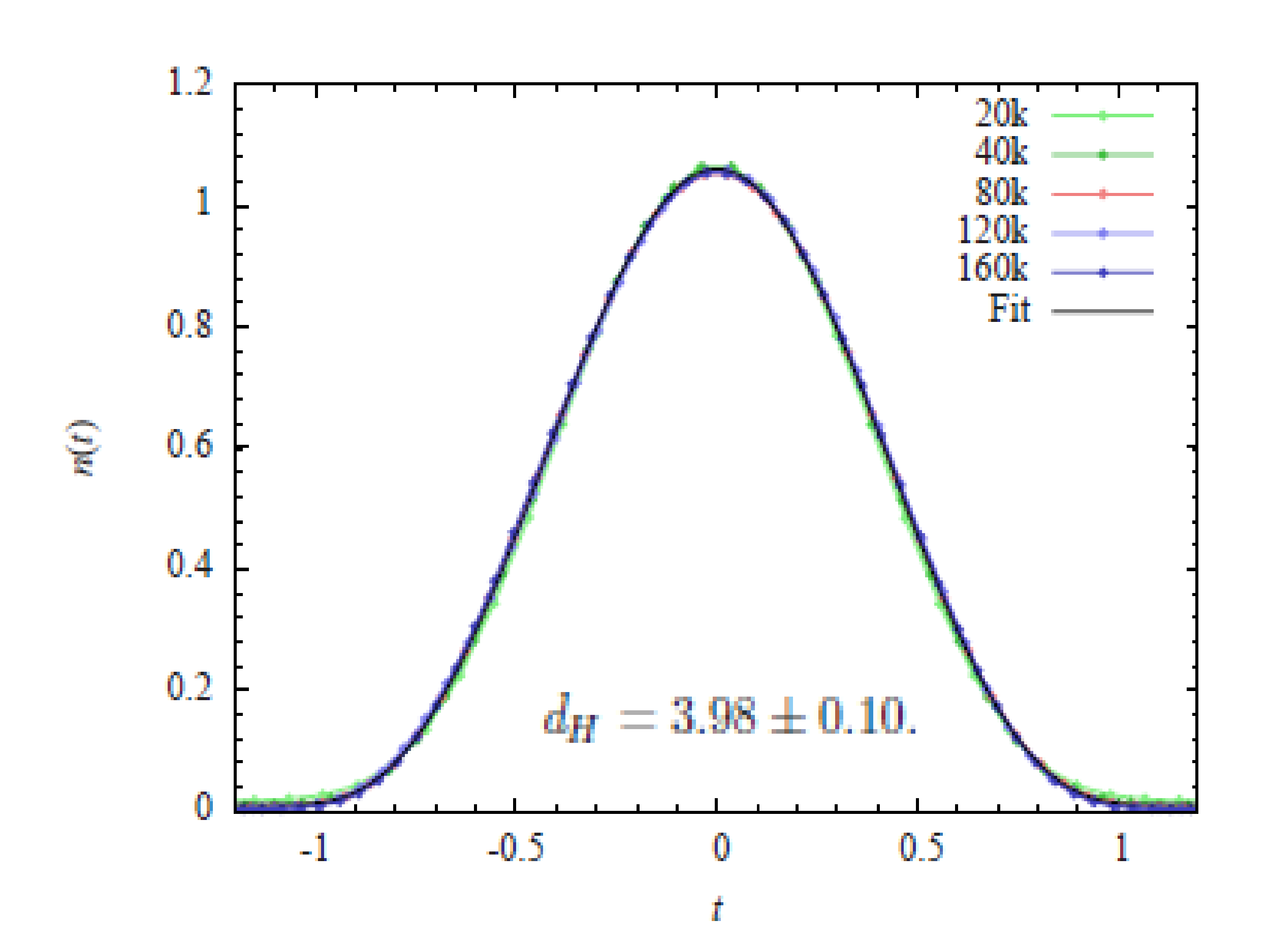}}}}
\caption{Scaling of the volume profiles.}
\label{fix}
\end{figure}
The plot illustrates the universality of the volume distribution for the rescaled
observable $n(\bar{t}) = N(t)/N_4^{1-1/d_H}$ plotted vs. $\bar{t}$. We expect the averaged scaled distribution
to be  volume-independent. This distribution can be interpreted as
a semi-classical limiting distribution of volume. Note that in our numerical experiments
it is obtained by integrating out all other degrees of freedom (details of the geometry)
except the spatial volume $N(t)$. The analysis shows that the averaged geometry
scales in a way consistent with dimension four. Although this result may appear trivial,
it is definitely not, since the distribution is obtained as the effect of a very delicate balance between
the entropy of configurations and the physical action. In  earlier studies, where causality was not imposed,
typical geometries dominating the quantum amplitude had either $d_H = 2$ (branched polymer  phase) or
$d_H = \infty$ (collapsed phase).

\begin{figure}[t]
\centerline{\scalebox{0.3}{\rotatebox{0}{\includegraphics{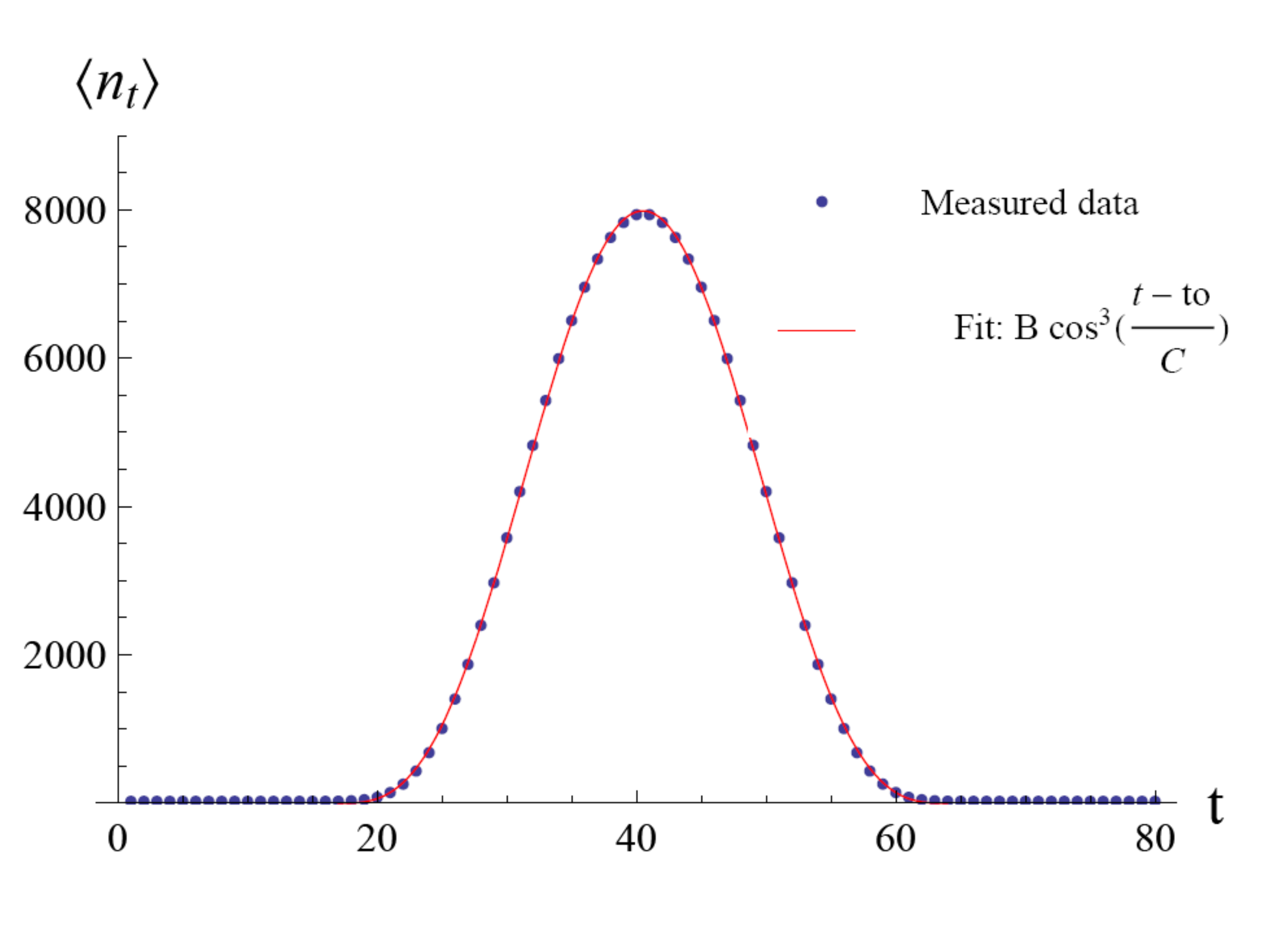}}}}
\caption{The averaged volume profile and the analytic formula used in the fit.}
\label{Fig03}
\end{figure}

We can analyze further the properties of the distribution and try to fit the limiting curve
by an analytic formula. 
The effect of the analysis is presented on the Fig.\ \ref{Fig03}, together with the fit. The analytic form
of the fit suggests that the observed geometry can be interpreted as the volume
dependence inside a four-dimensional ball, in this case the variable $t$ plays the role
of the azimuthal angle. This would indicate that in phase C we see the appearance of
a spherical four-dimensional de Sitter geometry.
 
The geometric properties  can be analyzed using other observables. A
useful example is that of the spectral dimension $d_S$. To measure this quantity
we analyze the return probability in the diffusion process on the geometry, as a function
of the diffusion time $\sigma$ \cite{ajl3}. If the geometry was regular we would expect
\begin{equation}
P(\sigma) \propto \frac{1}{\sigma^{d_S/2}},\quad d_S = -2\frac{d \log P(\sigma) }{d\log\sigma}
\label{e04}
\end{equation}
with a constant $d_S = d_H$. The figure shows the observed behaviour of $d_S$ obtained by averaging over many
starting points of the diffusion process and over many configurations.
\begin{figure}[h]
\centerline{\scalebox{0.35}{\rotatebox{0}{\includegraphics{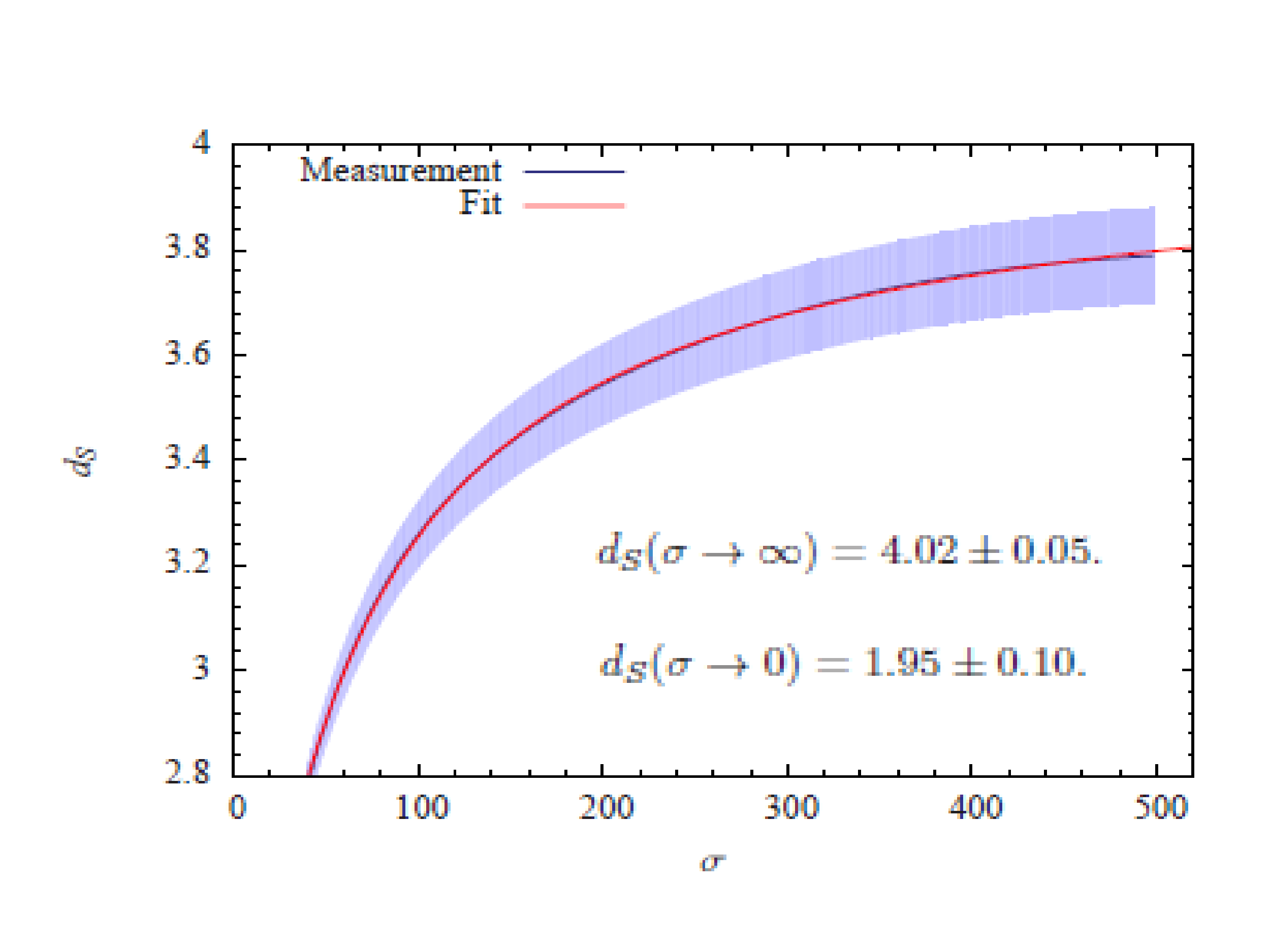}}}}
\caption{Scale dependence of $d_S$.}
\label{Fig04}
\end{figure}
The plot on Fig.\ \ref{Fig04} shows that $d_S$ is not a constant, but depends on $\sigma$ suggesting a scale dependence
of the effective geometry, ranging between two at short scales and four at large scales. This
illustrates the quantum character of geometry. A similar property was discovered in other
approaches to quantum gravity
(c.f. e.g. \cite{reutter1}).

The geometry presented above was measured at a particular point on the $\{\kappa_0,\Delta\}$ 
plane. When the values of the bare coupling constants are changed inside phase C, the qualitative behaviour remains the same, 
up to a finite change in the scale. We are particularly interested in the critical behaviour near
phase transitions.

The qualitative behaviour of the phase structure of CDT was found to have strong similarity to the phase structure
predicted by Ho\v{r}ava-Lifschitz gravity (\cite{horava} and \cite{jordan1}). The simplest check of the analogy was to measure
the order of the phase transitions, 
which we found to be first-order for the A-C transition and second order for the B-C transition \cite{jordan2}.

\subsection{Effective action in de Sitter phase}

The regular semi-classical distribution of spatial volume observed in phase C (de Sitter phase) suggests that it reproduces
 a saddle point of some effective action of the spatial volume $V(t)$ or, equivalently, of the scale factor $a(t) = V(t)^{1/3}$.
 A natural candidate for such
an action is the mini-superspace action
\begin{equation}
S = \frac{1}{24\pi G}\int~dt~\sqrt{g_{tt}}\left( \frac{g_{tt} V'(t)^2}{V(t)} +\tilde{\mu} V(t)^{1/3} - \tilde{\lambda} V(t)\right)
\label{e1}
\end{equation}
where $t$ takes the continuum value and $\tilde{\lambda}$ plays the role of Lagrange multiplier, necessary
to fix the total volume to some target volume $V$
\begin{equation}
V = \int~dt~ \sqrt{g_{tt}} V(t).
\label{e2}
\end{equation}
Here  $g_{tt}$ sets the scale in the time direction. In a discrete setup we may expect this action to take the discretized form
\begin{eqnarray}
S^{eff} &=& \frac{1}{\Gamma}\sum_t \left(\frac{(N_{t+1}-N_t)^2}{N_{t+1}+N_t} \right) \\ \nonumber
&+& \sum_t \left(\mu \left(\frac{N_{t+1}+N_t}{2}\right)^{1/3}-\lambda \frac{N_t +N_{t+1}}{2}\right) \\ \nonumber
&+&{\cal O}(N_t^{-1/3})
\label{seff}
\end{eqnarray}
although it could have a more complicated form.
Here $N_t \equiv N(t)$. 

Measuring the covariance matrix of volume fluctuations
around the semi-classical distribution and inverting this matrix we can determine the matrix of second derivatives of the effective action, assuming that higher-order
terms (higher than second order in fluctuations) can be neglected. 
This method was
successfully applied in \cite{agjl1}. Indeed the form (10) of the effective action
was confirmed, at least in the range of large volumes, permitting us to determine the physical parameters
$\Gamma$ and $\mu$. The parameter $\mu$ was particularly difficult to measure, since the corresponding term
in the action (after differentiating it twice) falls off very fast with the spatial volume. On the other
hand, small volumes can be expected to be (and are in fact) very sensitive to finite-size effects and
lattice artefacts.

The numerical experiment described above produces values of the physical parameters $\{\Gamma,\mu\}$
as functions of the bare couplings $\{\kappa_0,\Delta\}$. These physical parameters have a direct interpretation in terms 
of the gravitational constant $G$ (up to the dimensionful parameters $a_t$ and $a_s$). From a practical point of
view the determination of these parameters becomes more difficult near the phase transitions, where we observe
a critical slowing-down and large finite-size effects. As we observed, the neighbourhood of the phase transitions 
is particularly interesting from a physical point of view. Approaching these lines we see that
the parameter $1/\Gamma \to 0$ (or $\Gamma \to \infty$) \cite{we}, which can be interpreted as the limit where the
lattice spacing approaches zero. In this limit we may hope to observe genuine quantum effects of gravity.

\section{Transfer matrix}

The form of the effective action (10)  suggests a formal decomposition of the quantum amplitude (4)
in the simplified form
\begin{eqnarray}
&&G(N_0,N_T) = \\ \nonumber 
&&\hspace{-0.8cm}\sum_{N_1,N_2,\cdots,N_{T-1}}
\langle N_0|{\cM}|N_1\rangle \langle N_1|{\cM}|N_2\rangle\cdots\cdots
\langle N_{T-1} |{\cM}|N_T\rangle
\label{quant1}
\end{eqnarray}
where we have introduced the {\em effective} projection operators
\begin{eqnarray}
\rho(N)&\equiv& | N\rangle \langle N| = \sum_{T\in\cT_N} |T\rangle C(T) \langle T|, \\ \nonumber
\rho(N)\rho(N') &=& \delta_{NN'} \rho(N)
\label{qq}
\end{eqnarray}
on the space of states with a fixed volume $N$. The projection operators behave
as  genuine projection operators on a single state $|N\ra$.
They can be used 
to study the properties of the CDT geometry, assuming that the elements of the transfer
matrix $\sqrt{C(T_i)}\langle T_i|{\cM}|T_{i+1}\ra\sqrt{C(T_{i+1})} $ in (4) depend only on volume. 
We can check to what extent this is true.

In the proposed approach \cite{agsj}  we determine directly the elements 
$\langle N_i|{\cM}|N_{i+1}\rangle$ using numerical simulations of periodic
systems with very small time extent. The method is based on the observation that
terms in the sum (11) have the interpretation (up to a normalization) 
of the probability to measure a particular sequence of volumes. 
For a system with periodicity 2 and periodic boundary conditions
\begin{equation}
{\cal P}(N_1,N_2) \propto \langle N_1|{\cM}|N_2\rangle\langle N_2|{\cM}|N_1\rangle = \left(\langle N_1|{\cM}|N_2\rangle\right)^2
\label{3}
\end{equation}
By measuring the number of times a particular set $\{N_1,N_2\}$ appears in
the simulation we determine the matrix element $\langle N_1|{\cM}|N_2\rangle$.
In practice the method is more complicated, because we also want to study
a particular range of  $\{N_1,N_2\}$. Details of the method are explained
in \cite{agsj}. 
\begin{figure}[h]
\centerline{\scalebox{0.35}{\rotatebox{0}{\includegraphics{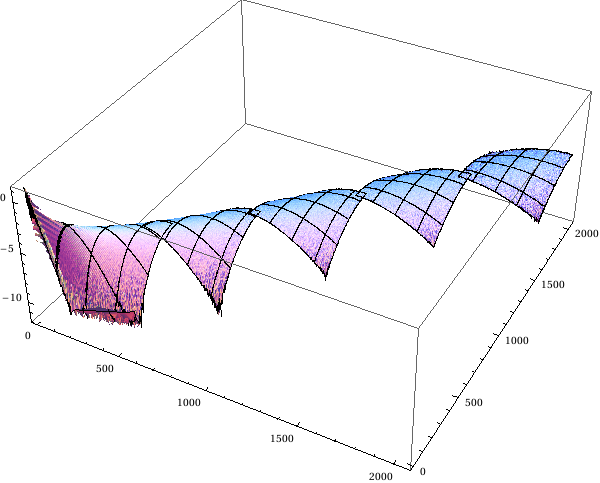}}}}
\caption{The logarithm of the transfer matrix elements plotted as a function of $N_1$ and $N_2$ for 
a sequence of ranges.}
\label{Fig06}
\end{figure}
On the Fig.\ \ref{Fig06} we show the logarithm of the transfer matrix, obtained by gluing together the results of measurements at
the neighbouring ranges of volume. On the plot we see the Gaussian behaviour of the off-diagonal
{\em kinetic} term and the diagonal {\em potential} terms. The kinetic term corresponds to the first
line in (10) and the potential term to the second line.
 We can easily measure the
parameters of the effective action and find  consistency with the values determined by the indirect
method described before. The advantage of the new approach is a much smaller numerical error
and at the same time a much shorter computer time needed to perform the measurements.

The presented plot corresponds to one particular point on the $\{\kappa_0,\Delta\}$ plane, well inside the de Sitter phase.
We are currently measuring the behaviour of physical parameters $\{\Gamma,\mu\}$ in the whole
range of the C (de Sitter) phase, in particular near the phase transitions. Preliminary results confirm that
$1/\Gamma \to 0$ at the A-C transition and inside the A phase and indicate that $1/\Gamma$ changes
sign at the B-C transition lines. Details of this behaviour are crucial to determine and understand the critical
behaviour near the phase transition and the critical scaling properties of the model.
Particularly interesting is the perspective to study the neighbourhood of the triple point, where the three phases meet.

\section{Conclusions}

The CDT model allows us to study properties of the lattice regularized quantum theory of geometry in a
Wick-rotated formulation (imaginary time). Obvious questions about the full properties of the model
under analytic continuation to real time remain open. Some features of the model are
however common to a formalism with real and imaginary time. One important property is the crucial role played by the entropy of configurations,
an aspect which is usually not appreciated in mini-superspace-type models.
Our approach is based on integrating out all degrees
of freedom, apart from a finite set and permits us to study the true effective model of the scale factor.

The new method to analyze the properties of the model, discussed in this article, gives rise to the hope that approaching
the boundaries of  de Sitter phase will be possible.


\begin{theacknowledgments}
JA and AG thank the Danish Research Council 
for financial support via the grant ``Quantum gravity and the role
of black holes'' and EU for
    support from the ERC-Advance grant 291092,
``Exploring the Quantum Universe'' (EQU). 
JJ acknowledges partial support of the International PhD Projects Programme of the Foundation
for Polish Science within the European Regional Development Fund of
the European Union, agreement no. MPD/2009/6.  JG-S acknowledges the Polish National Science Centre (NCN) support via the grant 2012/05/N/ST2/02698.

\end{theacknowledgments}

\end{document}